\documentclass[aps,twocolumn,showpacs,preprintnumbers,prb,floatfix]{revtex4}
\usepackage{graphicx}
\usepackage{bm}
\usepackage{color}

\begin{document}

\title{Total forces in the diffusion Monte Carlo method with nonlocal
pseudopotentials}

\author{A.~Badinski}
\author{R.~J.~Needs}

\affiliation{Theory of Condensed Matter Group, Cavendish Laboratory,
J.~J.~Thomson Avenue, Cambridge CB3 0HE, United Kingdom}

\date{\today}

\begin{abstract}
We report exact expressions for atomic forces in the diffusion Monte
Carlo (DMC) method when using nonlocal pseudopotentials.  We present
approximate schemes for estimating these expressions in both mixed and
pure DMC calculations, including the pseudopotential Pulay term
and the Pulay nodal term.
Harmonic vibrational frequencies and equilibrium bond lengths are
derived from the DMC forces and compared with those obtained from DMC
potential energy curves. Results for four small molecules show that
the equilibrium bond lengths obtained from our best force and energy
calculations differ by less than 0.002\,\AA.
\end{abstract}

%

\maketitle

\section{Introduction}

The diffusion Monte Carlo (DMC) method is the most accurate approach
available for calculating total ground-state energies of large
many-electron systems \citep{Foulkes}.  Energy derivatives, and in
particular atomic forces, are of great importance as they may be used
to relax atomic structures and perform molecular dynamics simulations.
It has, however, proven difficult to develop accurate and efficient
ways of calculating atomic forces within the DMC method.

The DMC method involves using the imaginary-time Schr\"odinger
equation to project away excited states of a many-electron
wavefunction so that the ground-state wavefunction $\Phi$ remains.
The fermionic symmetry is maintained by the fixed-node approximation
\citep{anderson} which constrains the nodal surface of $\Phi$ (the
hypersurface on which $\Phi$ is zero) to equal that of an
antisymmetric trial wavefunction $\Psi_{\mathrm{T}}$. The standard DMC
algorithm generates the ``mixed'' probability distribution
$\Psi_{\mathrm{T}}\Phi$ which can be used to calculate unbiased
expectation values (apart from the fixed-node error) of operators that
commute with the Hamiltonian, $\hat{H}$. If, however, an operator does
not commute with $\hat{H}$, the ``pure'' probability distribution
$\Phi\Phi$ is required which can be generated within the DMC method
using, for example, the future-walking algorithm \citep{Future}.

Within the Born-Oppenheimer approximation the atomic positions are
treated as parameters rather than dynamical variables, and the total
atomic force is defined as the negative energy gradient with respect
to the atomic position. In the mixed and pure DMC methods, the total
force estimators are constructed by taking the gradient of the (mixed
and pure) DMC energy. As in other electronic structure methods, these
estimators consist of contributions from the Hellmann-Feynman (HFT)
force \citep{Hellmann,Feynman} and from additional Pulay terms
\citep{Pulay,badinski2} which must be included to obtain unbiased
estimates of the forces. Although the estimators for the HFT force
have similar forms in the mixed and pure DMC methods, there are
significant differences between the Pulay terms.  The mixed DMC Pulay
term involves the derivative of the unknown DMC wavefunction $\Phi$,
which cannot be calculated straightforwardly. Reynolds $\textit{et
al.}$ \citep{Reynolds1,Reynolds2} suggested using the derivative of
$\Psi_{\mathrm{T}}$ instead and obtained a practical, although
approximate, scheme for estimating the mixed DMC Pulay term. The pure
DMC Pulay term involves an integral over the nodal surface
\citep{schautz1} which cannot be calculated in a standard DMC
calculation. A practical but approximate scheme for estimating this
nodal surface term has recently been developed \citep{badinski2}.

Although mixed DMC total forces have been investigated in studies of
small molecules \citep{Assaraf1,Assaraf2,Casalegno}, the pure DMC
nodal surface term \citep{chiesa,badinski1} has not been calculated
for real systems.  In a recent study of a model system without
electron-electron interactions, however, the nodal term was found to
be important \citep{badinski2}. Also, no previous study has directly
compared the performance of the mixed and pure DMC forces for real
systems.

It is well-known that the HFT estimator has an infinite variance when
the bare Coulomb potential is used to describe the electron-nucleus
interaction. Different routes have been proposed to address this
problem.  Assaraf $\textit{et al.}$ \citep{Assaraf1,Assaraf2,Assaraf3}
added a term to the HFT force which has a zero mean value but greatly
reduces the variance of the estimator. Chiesa $\textit{et al.}$
\citep{chiesa} developed a method to filter out the part of the
electron density that gives rise to the infinite variance.  Using soft
pseudopotentials also eliminates the infinite variance problem
\citep{badinski1}, and this method is used in the current work.

Pseudopotentials not only resolve the infinite variance issue when
calculating forces, they also remove the chemically inert core
electrons and their rapid spatial variations from the problem. This
greatly reduces the computational cost of a DMC calculation which
scales with the atomic number, $Z$, as $Z^{5.5}$. \citep{ceperley,ma}
However, the use of pseudopotentials introduces additional Pulay-like
terms in the HFT force estimator \citep{badinski1} which have been
neglected in previous calculations. In this work we have included
these pseudopotential Pulay terms.

We investigate the accuracy of the mixed and pure DMC force estimates
for the H$_{2}$, SiH, GeH and SiH$_{4}$ molecules.  Bond lengths and
harmonic vibrational frequencies are obtained from the forces and are
compared with those obtained from DMC energy calculations.  These
results are used to both evaluate the importance of the different
Pulay terms and to compare the performance of the mixed and pure DMC
force estimators.

This paper is organized as follows. In Sec.~2 we give exact and
approximate expressions for the forces under different pseudopotential
localization schemes. In Sec.~3 we describe our DMC calculations and
in Sec.~4 we present and discuss the molecular bond lengths and
vibrational frequencies obtained. We draw our conclusions in Sec.~5.

\section{Forces in the diffusion Monte Carlo method}

We write the valence Hamiltonian for a many-electron system as
\begin{equation}
\hat{H}=\hat{H}_{\mathrm{loc}}+\hat{W},
\label{eq:H_loc_W}
\end{equation}
where $\hat{H}_{\mathrm{loc}}$ consists of the kinetic energy, the
Coulomb interaction between the electrons and the local
pseudopotential, and $\hat{W}$ is the nonlocal pseudopotential
operator. Two different pseudopotential localization approximation
(PLA) schemes have been introduced to evaluate the nonlocal action of
$\hat{W}$ on the DMC wavefunction, $\Phi$. In these schemes $\hat{H}$
is replaced by an effective Hamiltonian \citep{hurley_PLA,Casula},
\begin{eqnarray}
\hat{H}_{A} & = & \hat{H}_{\mathrm{loc}}
+\frac{\hat{W}\Psi_{\mathrm{T}}}{\Psi_{\mathrm{T}}}, \label{eq:H_A}\\
\hat{H}_{B} & = & \hat{H}_{\mathrm{loc}}+\frac{\hat{W}^{+}
\Psi_{\mathrm{T}}}{\Psi_{\mathrm{T}}}+\hat{W}^{-}.
\label{eq:H_B}
\end{eqnarray}
The nonlocal pseudopotential operator $\hat{W}^{+}$ corresponds to all
positive matrix elements
$\left<\mathbf{r}'\right|\hat{W}^{+}\left|\mathbf{r}'\right>$, and
$\hat{W}^{-}$ corresponds to all negative matrix elements
\citep{Casula}, where $\mathbf{r}$ is the 3$N$-dimensional position
vector for the $N$ electron system and $N$ is the total number of
electrons. Following Ref.~\citep{badinski1}, these two approximations
are referred to as the full-PLA (FPLA) and semi-PLA (SPLA) when using
$\hat{H}_{A}$ or $\hat{H}_{B}$, respectively. The corresponding
fixed-node DMC ground-state wavefunctions are denoted by $\Phi_{A}$
and $\Phi_{B}$.

The DMC energy can be written in the form
\begin{equation}
E_{\mathrm{D}}=\frac{{\displaystyle{\int}\Phi\hat{H}\Psi\mathrm{d}V}}
{{\displaystyle{\int}\Phi\Psi\mathrm{d}V}},
\label{eq:E_d}
\end{equation}
which includes the mixed DMC ($\Psi=\Psi_{\mathrm{T}}$) and pure DMC
($\Psi=\Phi)$ estimates of the energy. In all later expressions,
$\Phi$ stands for either $\Phi_{A}$ or $\Phi_{B}$, and $\hat{H}$ for
either $\hat{H}_{A}$ or $\hat{H}_{B}$. Although the mixed and pure
estimates of $E_{\mathrm{D}}$ in Eq.~(\ref{eq:E_d}) are equivalent for
a given localization approximation, $E_{\mathrm{D}}$ may differ under
the two localization schemes.

We now consider a general parameter $\lambda$, e.g., a nuclear
coordinate, which is used to vary the Hamiltonian, and upon which both
the nodal surface (via $\Psi_{\mathrm{T}}$) and the DMC wavefunction
$\Phi$ depend. Taking the derivative of the DMC energy with respect to
$\lambda$ gives
\begin{eqnarray}
\frac{\mathrm{d}E_{\mathrm{D}}}{\mathrm{d}\lambda}
&=&\frac{{\displaystyle{\int}\left[\Phi\hat{H}'\Psi+\Phi\left(\hat{H}
-E_{\mathrm{D}}\right)\Psi'\right]\mathrm{d}V}}
{{\displaystyle{\int}\Phi\Psi\,\mathrm{d}V}}
\nonumber
\\
&&+\frac{{\displaystyle{\int}\left[\Phi'\left(\hat{H}
-E_{\mathrm{D}}\right)\Psi\right]\mathrm{d}V}}
{{\displaystyle{\int}\Phi\Psi\,\mathrm{d}V}},
\label{eq:F_general}
\end{eqnarray}
for both the mixed and pure DMC methods. We use the notation
$\alpha'=\frac{\mathrm{d}\alpha}{\mathrm{d}\lambda}$ where $\alpha$
can be a function or an operator. The first term in
Eq.~(\ref{eq:F_general}) is the HFT force \citep{Hellmann,Feynman} and
the others are Pulay terms \citep{Pulay}.

\subsection{Mixed DMC forces}

The total force in the mixed DMC method,
$F_{\mathrm{mix}}^{\mathrm{tot}}$, is obtained by setting
$\Psi=\Psi_{\mathrm{T}}$ in Eq.~(\ref{eq:F_general}). After some
rearrangements, we arrive at
\begin{equation}
F_{\mathrm{mix}}^{\mathrm{tot}}=F_{\mathrm{mix}}^{\mathrm{HFT}}
+F_{\mathrm{mix}}^{\mathrm{P}}+F_{\mathrm{mix}}^{\mathrm{V}}
+F_{\mathrm{mix}}^{\mathrm{N}},
\label{eq:F_tot,mix}
\end{equation}
with
\begin{eqnarray}
F_{\mathrm{mix}}^{\mathrm{HFT}} & = & 
-\frac{{\displaystyle{\int}\Phi\Psi_{\mathrm{T}}\left(\frac{\hat{W}'
\Psi_{\mathrm{T}}}{\Psi_{\mathrm{T}}}\right)\mathrm{d}V}}
{{\displaystyle{\int}\Phi\Psi_{\mathrm{T}}\,\mathrm{d}V}}
\nonumber
-\frac{{\displaystyle{\int}\Phi\Psi_{\mathrm{T}}V_{\mathrm{loc}}'\mathrm{d}V}}
{{\displaystyle{\int}\Phi\Psi_{\mathrm{T}}\,\mathrm{d}V}}
\nonumber \\
&&
+ Z_\alpha \sum_{\beta\,(\beta \ne \alpha)} Z_{\beta} \frac{\mathbf{R}_{\alpha}
-\mathbf{R}_{\beta}}{|\mathbf{R}_{\alpha}-\mathbf{R}_{\beta}|^3}
\label{eq:F_HFT,mix}
\\
F_{\mathrm{mix}}^{\mathrm{P}} & = & 
-\frac{{\displaystyle{\int}\Phi\Psi_{\mathrm{T}}\left[\frac{\hat{W}
\Psi_{\mathrm{T}}'}
{\Psi_{\mathrm{T}}}-\left(\frac{\hat{W}\Psi_{\mathrm{T}}}
{\Psi_{\mathrm{T}}}\right)\frac{\Psi_{\mathrm{T}}'}
{\Psi_{\mathrm{T}}}\right]\mathrm{d}V}}
{{\displaystyle{\int}\Phi\Psi_{\mathrm{T}}\,\mathrm{d}V}}
\label{eq:F_PPT,mix}\\
F_{\mathrm{mix}}^{\mathrm{V}} & = & 
-\frac{{\displaystyle{\int}\Phi\Psi_{\mathrm{T}}\left[\frac{\Phi'}{\Phi}
\frac{\left(\hat{H}-E_{\mathrm{D}}\right)\Psi_{\mathrm{T}}}
{\Psi_{\mathrm{T}}}\right]\mathrm{d}V}}
{{\displaystyle{\int}\Phi\Psi_{\mathrm{T}}\,\mathrm{d}V}}
\label{eq:F_V}\\
F_{\mathrm{mix}}^{\mathrm{N}} & = & 
-\frac{{\displaystyle{\int}\Phi\Psi_{\mathrm{T}}\left[\frac{\left(\hat{H}
-E_{\mathrm{D}}\right)\Psi_{\mathrm{T}}'}{\Psi_{\mathrm{T}}}
\right]\mathrm{d}V}}{{\displaystyle{\int}\Phi\Psi_{\mathrm{T}}\,\mathrm{d}V}}.
\label{eq:F_N,mix}
\end{eqnarray}
$F_{\mathrm{mix}}^{\mathrm{HFT}}$ is the mixed DMC HFT force and the
other expressions are Pulay terms. The HFT force in
Eq.~(\ref{eq:F_HFT,mix}) contains two contributions from the
pseudopotential, one from its local part $V_{\mathrm{loc}}$ and one
from its nonlocal part $\hat{W}$, and a third contribution from the
nucleus-nucleus interaction. In this nucleus-nucleus interaction term,
$\mathbf{R}_{\alpha}$ represents the $3$-dimensional position vector
of the $\alpha$th nucleus, and $Z_{\alpha}$ is the associated charge.
The three Pulay terms in Eqs.~(\ref{eq:F_PPT,mix})-(\ref{eq:F_N,mix})
are identified as follows: $F_{\mathrm{mix}}^{\mathrm{P}}$ results
from the PLA and is therefore called the pseudopotential Pulay term,
$F_{\mathrm{mix}}^{\mathrm{V}}$ is the volume term, and
$F_{\mathrm{mix}}^{\mathrm{N}}$ is called the mixed DMC nodal term
since it can be written as an integral over the nodal surface
\citep{badinski2}. Note that all terms in
Eqs.~(\ref{eq:F_HFT,mix})-(\ref{eq:F_N,mix}) take the same form under
both localization schemes, the only difference is the distribution
($\Psi_{\mathrm{T}}\Phi_A$ or $\Psi_{\mathrm{T}}\Phi_B$) used to
evaluate the expectation values.  A simple way to understand this is
to note that $\hat{H}$ always acts on the trial wavefunction
$\Psi_{\mathrm{T}}$ and
$\hat{H}_A\Psi_{\mathrm{T}}=\hat{H}_B\Psi_{\mathrm{T}}$.

In mixed DMC simulations, it is straightforward to evaluate the
contributions to the force except for the volume term
$F_{\mathrm{mix}}^{\mathrm{V}}$, because it depends on the derivative
of the DMC wavefunction, $\Phi'$.  Since it is unclear how to evaluate
$\Phi'$ in mixed DMC calculations, we use the Reynolds' approximation
\citep{Reynolds1,Reynolds2},
\begin{equation}
\frac{\Phi'}{\Phi}\simeq\frac{\Psi_{\mathrm{T}}'}{\Psi_{\mathrm{T}}},
\label{eq:Reynolds_approx}
\end{equation}
which is exact on the nodal surface (see Eqs.~(4) and (16) of
Ref.~\citep{badinski2}) but introduces an error of first order in
($\Psi_{\mathrm{T}}-\Phi$) away from the nodal surface.  

\subsection{Pure DMC forces }

The total force in the pure DMC method,
$F_{\mathrm{pure}}^{\mathrm{tot}}$, is obtained by setting $\Psi=\Phi$
in Eq.~(\ref{eq:F_general}). After some manipulations, we obtain
\begin{equation}
F_{\mathrm{pure}}^{\mathrm{tot}}=F_{\mathrm{pure}}^{\mathrm{HFT}}
+F_{\mathrm{pure}}^{\mathrm{P}}+F_{\mathrm{pure}}^{\mathrm{N}},
\label{eq:F_tot_pure}
\end{equation}
with
\begin{eqnarray}
F_{\mathrm{pure}}^{\mathrm{HFT}} & = & \left\{ \begin{array}{c}
-\frac{{\displaystyle{\int}\Phi_{A}\Phi_{A}\left(\frac{\hat{W}'\Psi_{\mathrm{T}}}
{\Psi_{\mathrm{T}}}\right)\mathrm{d}V}}{{\displaystyle{\int}\Phi_{\mathrm{A}}
\Phi_{\mathrm{A}}\,\mathrm{d}V}}\,\,\,\,\,\,\,\,\,\,\,\,\,\,\,\,\,\,\,\,\,\\
-\frac{{\displaystyle{\int}\Phi_{B}\Phi_{B}\left(\frac{(\hat{W}^{+})'\Psi_{\mathrm{T}}}
{\Psi_{\mathrm{T}}}+\frac{(\hat{W}^{-})'\Phi_{\mathrm{B}}}
{\Phi_{\mathrm{B}}}\right)\mathrm{d}V}}{{\displaystyle{\int}
\Phi_{B}\Phi_{B}\,\mathrm{d}V}}
\end{array}\right\}
\nonumber\\ 
&& -\frac{{\displaystyle{\int}\Phi\Phi V'_{\mathrm{loc}}\mathrm{d}V}}
{{\displaystyle{\int}\Phi\Phi\,\mathrm{d}V}}
 + Z_\alpha \sum_{\beta\,(\alpha \ne \beta)} Z_{\beta} \frac{\mathbf{R}_{\alpha}
-\mathbf{R}_{\beta}}{|\mathbf{R}_{\alpha}-\mathbf{R}_{\beta}|^3}
\label{eq:F_HFT_pure}
\\
F_{\mathrm{pure}}^{\mathrm{P}} & = & \left\{ \begin{array}{c}
-\frac{{\displaystyle{\int}\Phi_{A}\Phi_{A}\left[\frac{\hat{W}\Psi_{\mathrm{T}}'}
{\Psi_{\mathrm{T}}}-\left(\frac{\hat{W}\Psi_{\mathrm{T}}}
{\Psi_{\mathrm{T}}}\right)\frac{\Psi_{\mathrm{T}}'}
{\Psi_{\mathrm{T}}}\right]\mathrm{d}V}}{{\displaystyle{\int}
\Phi_{A}\Phi_{A}\,\mathrm{d}V}} \,\,\,\,\,\,\, \\
-\frac{{\displaystyle{\int}\Phi_{B}\Phi_{B}\left[\frac{\hat{W}^{+}\Psi_{\mathrm{T}}'}
{\Psi_{\mathrm{T}}}-\left(\frac{\hat{W}^{+}\Psi_{\mathrm{T}}}
{\Psi_{\mathrm{T}}}\right)\frac{\Psi_{\mathrm{T}}'}{\Psi_{\mathrm{T}}}
\right]\mathrm{d}V}}{{\displaystyle{\int}\Phi_{B}\Phi_{B}\,\mathrm{d}V}}
\end{array}\right.
\label{eq:F_PPT_pure}
\\
F_{\mathrm{pure}}^{\mathrm{N}} & = & \frac{1}{2}\frac{{\displaystyle{\int}_{\Gamma}|
\nabla_{\mathbf{r}}\Phi|\Phi'\, \mathrm{d}S}}{{\displaystyle{\int}
\Phi\Phi \mathrm{d}V}}.
\label{eq:F_N_pure}
\end{eqnarray}
$F_{\mathrm{pure}}^{\mathrm{HFT}}$ is the pure DMC HFT force,
$F_{\mathrm{pure}}^{\mathrm{P}}$ is the pure DMC pseudopotential Pulay
term, and the pure DMC nodal term $F_{\mathrm{pure}}^{\mathrm{N}}$ is
an integral over the nodal surface $\Gamma$ (defined by
$\Psi_{\mathrm{T}}=0$).
Where terms appear in braces, the upper one refers to the FPLA and the
lower to the SPLA.
The form of the nodal term in Eq.~(\ref{eq:F_N_pure}) is independent
of the localization scheme. The nodal term involves the gradient
$\nabla_{\mathbf{r}}\Phi$ evaluated at the nodal surface
$\Gamma$. Ref.~\citep{badinski2} shows that this gradient is defined
as its limit when approaching the nodal surface from within a nodal
pocket (where $\Phi$ is nonzero). The derivation of the nodal term
from Eq.~(\ref{eq:F_general}) can be found in
Refs.~\citep{schautz1,badinski2}.

Although the HFT force $F_{\mathrm{pure}}^{\mathrm{HFT}}$ under the
FPLA can be calculated in the pure DMC method, it is not
straightforward to evaluate the action of the nonlocal operator
$(\hat{W}^{-})'$ on the unknown DMC wavefunction $\Phi_{B}$ in
$F_{\mathrm{pure}}^{\mathrm{HFT}}$ under the SPLA scheme. Therefore,
the following localization approximation,
\begin{equation}
\frac{(\hat{W}^{-})'\Phi_{B}}{\Phi_{B}}\simeq\frac{(\hat{W}^{-})
'\Psi_{\mathrm{T}}}{\Psi_{\mathrm{T}}},\label{eq:approx_under_SPLA}
\end{equation}
is used in the evaluation of $F_{\mathrm{pure}}^{\mathrm{HFT}}$ under
the SPLA scheme which introduces an error of first order in
$(\Psi_{\mathrm{T}}-\Phi)$.

Another complication arises with the pure DMC nodal term
$F_{\mathrm{pure}}^{\mathrm{N}}$ in Eq.~(\ref{eq:F_N_pure}) because it
involves the evaluation of an integral over the nodal surface. It is
unclear how to evaluate such an integral in a standard DMC simulation.
The following relationship suggested in Ref.~\citep{badinski2},
\begin{equation}
F_{\mathrm{pure}}^{\mathrm{N}}=2\, F_{\mathrm{mix}}^{\mathrm{N}}
+\mathcal{O}[(\Psi_{\mathrm{T}}-\Phi)^{2}],
\label{eq:F_N_pure_approx}
\end{equation}
allows the approximate evaluation of $F_{\mathrm{pure}}^{\mathrm{N}}$
as twice its mixed DMC counterpart while introducing an error of
second order in $(\Psi_{\mathrm{T}}-\Phi)$.
$F_{\mathrm{mix}}^{\mathrm{N}}$ may be evaluated in a standard DMC
simulation using the volume integral representation of
Eq.~(\ref{eq:F_N,mix}).  Equation~(\ref{eq:F_N_pure_approx}) is an
application of the standard extrapolation technique \citep{Foulkes},
as in this case the variational estimate of the nodal term is zero
\citep{badinski2}.

It is worth stating that the pseudopotential Pulay terms in both mixed
and pure DMC simulations vanish when $\Psi_{\mathrm{T}}$ equals
$\Phi$, which follows by inspection. Also, the mixed and pure DMC
nodal terms vanish when the nodal surface of $\Psi_{\mathrm{T}}$ is
exact. The proof of this statement is analogous to the one presented
in Appendix C of Ref.~\citep{badinski2}.

\subsection{Total versus partial derivatives }\label{sub:tot_vs_partial}

Since the atomic force equals the negative total derivative of the DMC
energy with respect to a nucleus position $\lambda$, all previous
expressions involve total derivatives, in particular
$\Psi_{\mathrm{T}}'$. Calculating the total derivatives involves
knowledge of how the wavefunction changes with $\lambda$. In the
variational Monte Carlo (VMC) method, all parameters $\{c_{i}\}$ that
describe the wavefunction $\Psi_{\mathrm{T}}$ can in principle be
uniquely specified, e.g., by minimizing the variational energy. The
specification of all $\{c_{i}\}$ parameters does not, however,
uniquely determine the derivative of $\Psi_{\mathrm{T}}$ with respect
to $\lambda$. In standard quantum chemistry methods, the derivatives
of the $c_{i}$ with respect to $\lambda$ are typically obtained by
second-order perturbation theory \citep{coupled_HF,per}.
Unfortunately, such an approach is not straightforward in VMC and DMC
methods.

In this work, we follow a different route and approximate all total
derivatives by their partial derivatives, which introduces an error of
first order in $(\Psi_{\mathrm{T}}-\Phi)$. We expect, however, this
approximation to be rather accurate for the following reason: taking
the total derivative of the DMC energy with respect to $\lambda$ gives
\begin{equation}
\frac{dE_{\mathrm{D}}}{d\lambda}=\frac{\partial E_{\mathrm{D}}}
{\partial\lambda}+\sum_{i}\frac{\partial E_{\mathrm{D}}}
{\partial c_{i}}\frac{dc_{i}}{d\lambda},
\label{eq:tot_vs_part}
\end{equation}
where the $c_{i}$ are the parameters in $\Psi_{\mathrm{T}}$ and the
Hamiltonian. The sum in Eq.~(\ref{eq:tot_vs_part}) stems from the
implicit dependencies of the parameters $c_{i}$ on $\lambda$. This sum
is neglected when all total derivatives are replaced with partial
derivatives in all previous force expressions.  Since the DMC energy
is approximately minimized with respect to the $c_{i}$, we expect that
the parameters $c_{i}$ have little effect on the DMC energy, i.e.,
$\frac{\partial E_{\mathrm{D}}}{\partial c_{i}}$ is small. Therefore,
neglecting the sum in Eq.~(\ref{eq:tot_vs_part}), or equivalently
replacing all total derivatives with partials in our previous
expressions, is expected to be a good approximation.

\section{Details of QMC calculations}

\begin{table*}\begin{centering}\begin{tabular}{lllllll}
\hline\hline 
&  BL ($\mathrm{\AA}$)&  Basis&  $E_{\mathrm{HF}}$(Ha)& 
   $E_{\mathrm{VMC}}$(Ha)&  $E_{\mathrm{DMC}}$(Ha)&  $E_{\mathrm{C}}$\\
\hline
H$_{2}$  &  0.740&  large&  -1.13367&  -1.17399(1)&  -1.17452(1)&  98.7$\,$\%\\
SiH      &  1.520&  large&  -4.26235&  -4.36967(4)&  -4.37719(2)&  93.4$\,$\%\\
GeH      &  1.600&  large&  -4.24392&  -4.34377(1)&  -4.35143(2)&  92.9$\,$\%\\
SiH$_{4}$&  1.480&  large&  -6.08924&  -6.27247(3)&  -6.27927(7)&  96.4$\,$\%\\
\hline 
SiH      &  1.520&  small&  -4.24689&  -4.36563(6)&  -4.37611(2)&  91.9$\,$\%\\
GeH      &  1.600&  small&  -4.23275&  -4.34156(2)&  -4.34928(2)&  93.4$\,$\%\\
\hline\hline 
\end{tabular}
\par\end{centering}
\caption{\label{tab:energies} Hartree-Fock (HF), VMC and DMC energies
(Ha) for four molecules. The first column specifies the bond length
(BL) used, the second column gives the basis set, and the last column
states the percentage of the DMC correlation energy retrieved within
the VMC method, $E_{\mathrm{C}}$, as a measure of the accuracy of the
Jastrow factor as defined in Ref.~\citep{badinski1}. The error bars in
$E_{\mathrm{C}}$ are smaller than 0.1$\,$\%.}
\end{table*}

We use a trial wavefunction of the standard single-determinant
Slater-Jastrow\citep{Foulkes} form. The orbitals forming the
Slater-determinants are obtained from Hartree-Fock calculations using
the $\textsc{gamess-us}$ \citep{gamess} code with atomic-centered
Gaussian basis sets. For all molecules, we use a basis set of
sextuple-$\zeta$ quality (without $f$ and $g$ functions but with four
additional diffuse $p$ and $d$ functions).  To study the influence of
the basis set we also use a smaller Gaussian basis set for the SiH and
GeH molecules with only five $s$ and two $p$-functions so that the
trial wavefunctions $\Psi_{\mathrm{T}}$ and the nodal surface are less
accurate. We refer to these two basis sets as $\textit{large}$ and
$\textit{small}$.

Table \ref{tab:energies} shows that, when using the small instead of
the large basis set, the DMC total energies increase by 1.01(3)\,mHa
and 2.15(3)\,mHa for the SiH and GeH molecules, respectively.

We use Jastrow factors consisting of electron-electron,
electron-nucleus, and electron-electron-nucleus terms, which are
expanded in natural power series \citep{drummond_jastrow}. The
wavefunction for H$_{2}$ has 87 variable parameters, while those for
the other molecules have 157. The parameters in the Jastrow factors
are optimized by first minimizing the variance of the local energy
\citep{drummond_variance}, and subsequently minimizing the energy
\citep{toulouse,brown} while keeping the cutoff parameters of the
Jastrow factor fixed \citep{drummond_jastrow}.  We use Dirac-Fock
averaged relativistic effective pseudopotentials \citep{trail2,trail1}
which can be obtained online \citep{casino_page}. The future-walking
method \citep{Future} is used to sample the pure estimates and all DMC
calculations are performed using the $\textsc{casino}$ code
\citep{casino} version 2.1.

We use the analytic expressions derived in Ref.~\citep{badinski1} for
evaluating the HFT force. The Pulay terms may also be evaluated using
analytic expressions, but to make the code more easily adaptable to
other trial wavefunction forms we use a finite-difference approach.
This introduces an error which is linear in the infinitesimal nuclear
displacement, $\Delta$.
We find that $\Delta \approx 10^{-7}$~\AA\ minimizes the resulting
error in the Pulay terms which is about seven orders of magnitude
smaller than the estimated values of the total forces.

\begin{figure}[t]
\begin{centering}
\includegraphics[scale=0.55]{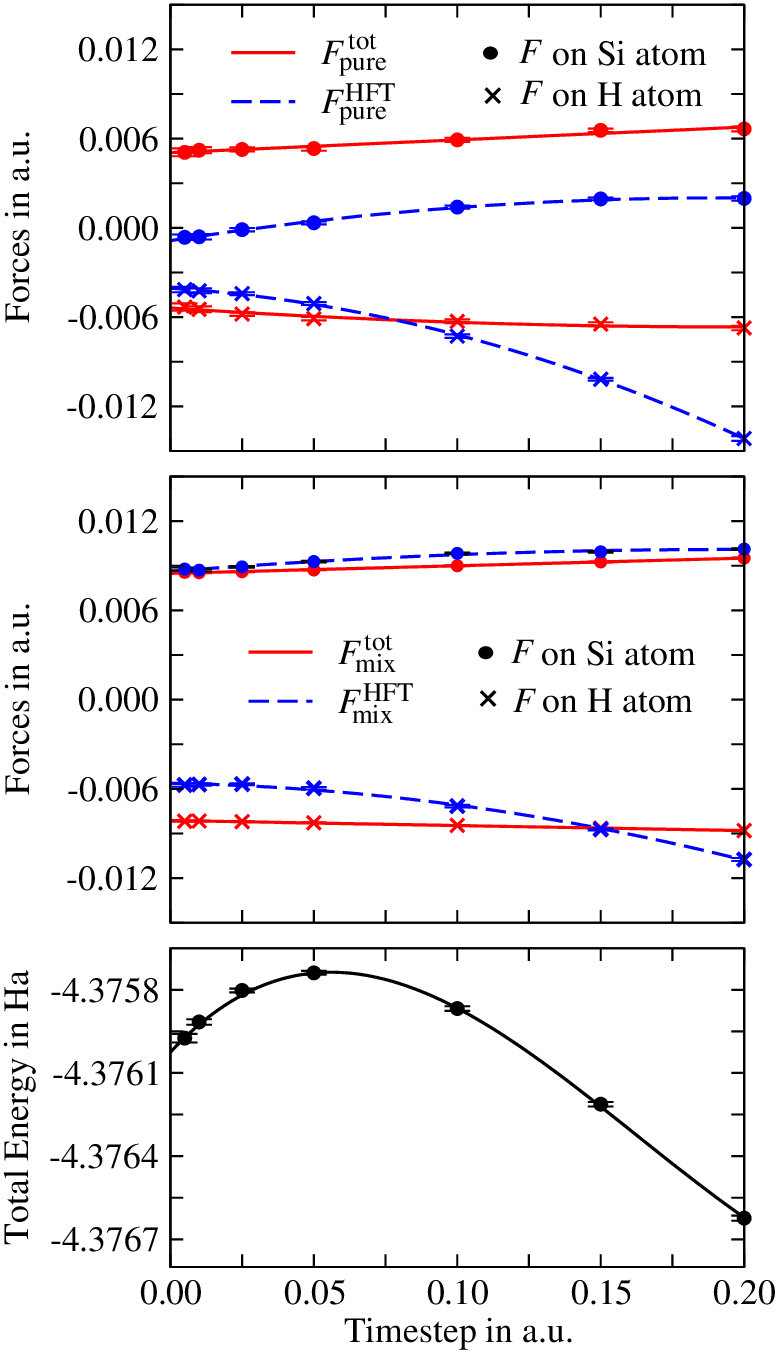}
\par\end{centering}
\caption{\label{fig:timestep} (Color online) Investigation of the
timestep behaviour of DMC forces (a.u.) and DMC energies (Ha) for the
SiH molecule when using the small basis set. Upper graph: the pure DMC
forces $F_{\mathrm{pure}}^{\mathrm{tot}}$ and
$F_{\mathrm{pure}}^{\mathrm{HFT}}$ as a function of the DMC timestep,
circles indicate forces on the Si atom and crosses indicate forces on
the H atom. A quadratic form is fitted to these forces. The standard
errors of these forces are indicated unless they are smaller than the
line width of the fitted functions. Middle graph: the same information
as the upper graph for mixed DMC calculations. Lower graph: the DMC
energy as a function of timestep for comparison, a cubic form is
fitted to the energies.}
\end{figure}

DMC calculations suffer from systematic errors arising from the
short-time approximation to the Green's function, which we have
carefully investigated. We find that the forces calculated with DMC
timesteps of 0.01 and 0.005\,a.u.\ agree with forces obtained from
extrapolations to zero timestep within one standard error of less than
0.0005\,a.u.\ Figure~\ref{fig:timestep} shows such an investigation
for the SiH molecule using the small basis set where the forces acting
on the Si and H atoms are plotted as a function of timestep.  We
therefore use a timestep of 0.005\,a.u.\ for all our DMC calculations,
to avoid the necessity for repeated extrapolation to zero timestep.
It is worth noting that, for the systems studied here, the timestep
errors in the HFT and Pulay forces tend to cancel one another.  This
can be seen for the SiH molecule in Figure~\ref{fig:timestep} when
comparing the solid lines (total forces) with the dashed lines (HFT
forces).
\begin{figure}
\begin{centering}
\includegraphics[scale=0.55]{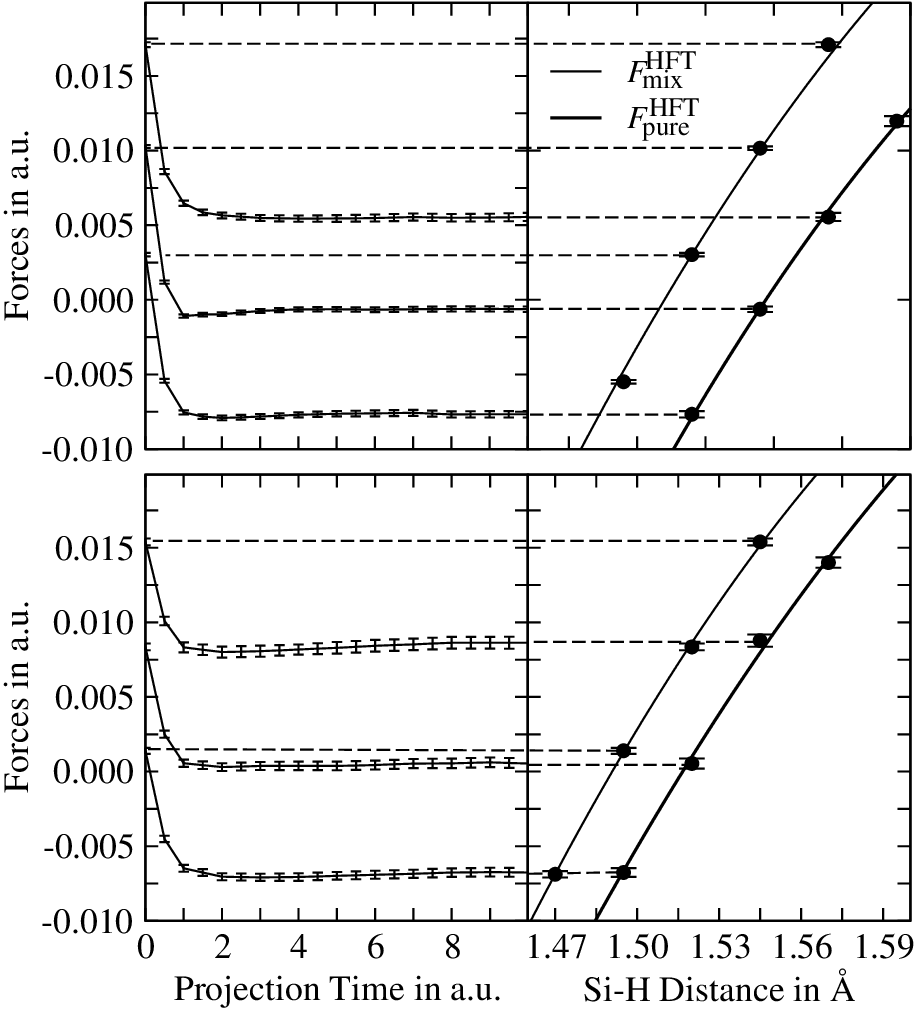}
\par\end{centering}
\caption{\label{fig:proj_time} Dependence of the future-walking pure
DMC HFT force estimates (a.u.) on the future-walking projection time
(a.u.).  Upper right graph: the HFT force on the Si atom for the SiH
molecule as calculated within the mixed DMC and future-walking pure
DMC method for different bond lengths using the small basis set. A
Morse potential is fitted to these forces. Upper left graph:
future-walking pure DMC forces calculated for different bond lengths
plotted against the future-walking projection time.  The mixed DMC
force corresponds to zero future-walking projection time. Lower
graphs: the same information as the upper graphs for the large basis
set. When using the small or large basis set, we see that the
future-walking projection time of 10\,a.u.\ is long enough to obtain
accurate pure DMC estimates.}
\end{figure}

For the future-walking pure DMC estimates to be accurate, an infinite
future-walking projection time is in principle required. However, we
found a projection time of 10\,a.u.\ to be sufficient in our
calculations as no significant changes in the estimates were found
when using longer projection times. For example, Figure
\ref{fig:proj_time} shows the convergence of the pure HFT forces with
respect to the projection time for the SiH molecule and both basis
sets.

\begin{table*}\begin{centering}\begin{tabular}{llllllll}
\hline\hline 
Form& Basis& $a^{E}$& $a_{\mathrm{mix}}^{F_{\mathrm{tot}}}$&
$a_{\mathrm{pure}}^{F\mathrm{_{tot}}}$& $\omega^{E}$&
$\omega_{\mathrm{mix}}^{F_{\mathrm{tot}}}$&
$\omega_{\mathrm{pure}}^{F_{\mathrm{tot}}}$\\
\hline
P(3/2)& small& 1.5242(7)& 1.5138(1)& 1.5259(8)& 2000(18)& 2096(4)& 2050(14)\\
P(4/3)& small& 1.5238(7)& 1.5138(1)& 1.5261(7)& 1983(55)& 2095(5)& 2018(28)\\
Morse & small& 1.5242(6)& 1.5141(1)& 1.5259(6)& 1992(12)& 2084(2)& 2045(11)\\
\hline 
P(3/2)& large& 1.5195(8)& 1.5105(2)& 1.5175(11)&2069(18)& 2089(4)& 2061(18)\\
P(4/3)& large& 1.5194(8)& 1.5107(2)& 1.5177(11)&2104(56)& 2078(6)& 2046(38)\\
Morse & large& 1.5195(7)& 1.5107(1)& 1.5173(9)& 2049(13)& 2080(2)& 2052(11)\\
\hline 
Expt. & & 1.520& 1.520& 1.520& 2042& 2042& 2042\\
\hline\hline 
\end{tabular}
\par\end{centering}
\caption{\label{tab:fit}Bond lengths, $a$ ($\mathrm{\AA}$), and
harmonic vibrational frequencies, $\omega$ (cm$^{-1}$), for the SiH
molecule derived from different functional forms previously fitted to
the DMC energies and forces. Three fitting forms are used as indicated
in the first column, P(3/2) indicates that a cubic polynomial is
fitted to the energies and a quadratic one to the forces, and
similarly for P(4/3), and Morse indicates that a Morse potential is
fitted to the energy and its derivative to the forces. The results are
compared with the experimental values taken from Ref.~\citep{NIST}.}
\end{table*}

To determine the equilibrium bond lengths, we calculate the forces at
0\,\%, $\pm$1.5\,\%, $\pm$3\,\%, and $\pm$4.5\,\% around the
experimental bond lengths. We then fit the derivative of the Morse
potential \citep{morse} with three free parameters to our force data
and locate the zero-force geometry and the harmonic vibrational
frequency. For all molecules, we compare results derived from the
Morse potential with those obtained from quadratic and cubic fitting
forms. We find that the influence of the different fitting forms on
the derived bond lengths and frequencies is much less than one
standard error except for a few cases when the forces are obtained
from the mixed DMC total force estimator.
There, the statistical error is much reduced and the influence of the
fitting form on the derived bond lengths and frequencies is sometimes
as large as two standard errors. Table \ref{tab:fit}, for example,
presents results obtained from the different fitting forms as
calculated for the SiH molecule with both basis sets. It is not
obvious which fitting form is best suited to describe our fitted
data. However, the Morse potential gives the smallest statistical
error bars for all derived bond lengths and frequencies, which
suggests that the Morse potential may indeed give the best description
of our data.
We also obtain bond lengths and frequencies from a Morse potential
fitted to the calculated DMC energies.

The statistical error bars in the calculated bond lengths and
frequencies are determined using a statistical method. For
convenience, the DMC forces and energies evaluated at the seven
different bond lengths are called a \textit{set of data}. For each set
of data, statistical noise is added where the noise is obtained from a
Gaussian distribution with a variance specified by one standard error
of the DMC force and energy estimates. 10$^5$ different sets of data
are generated in this manner and a Morse potential is fitted to these
sets. The bond lengths and frequencies are obtained for all sets, they
are averaged, and their statistical error bars are
determined. Throughout this work, we use the convention that one
standard error corresponds to a one-sigma confidence interval.

\begin{figure}
\begin{centering}
\includegraphics[scale=0.53]{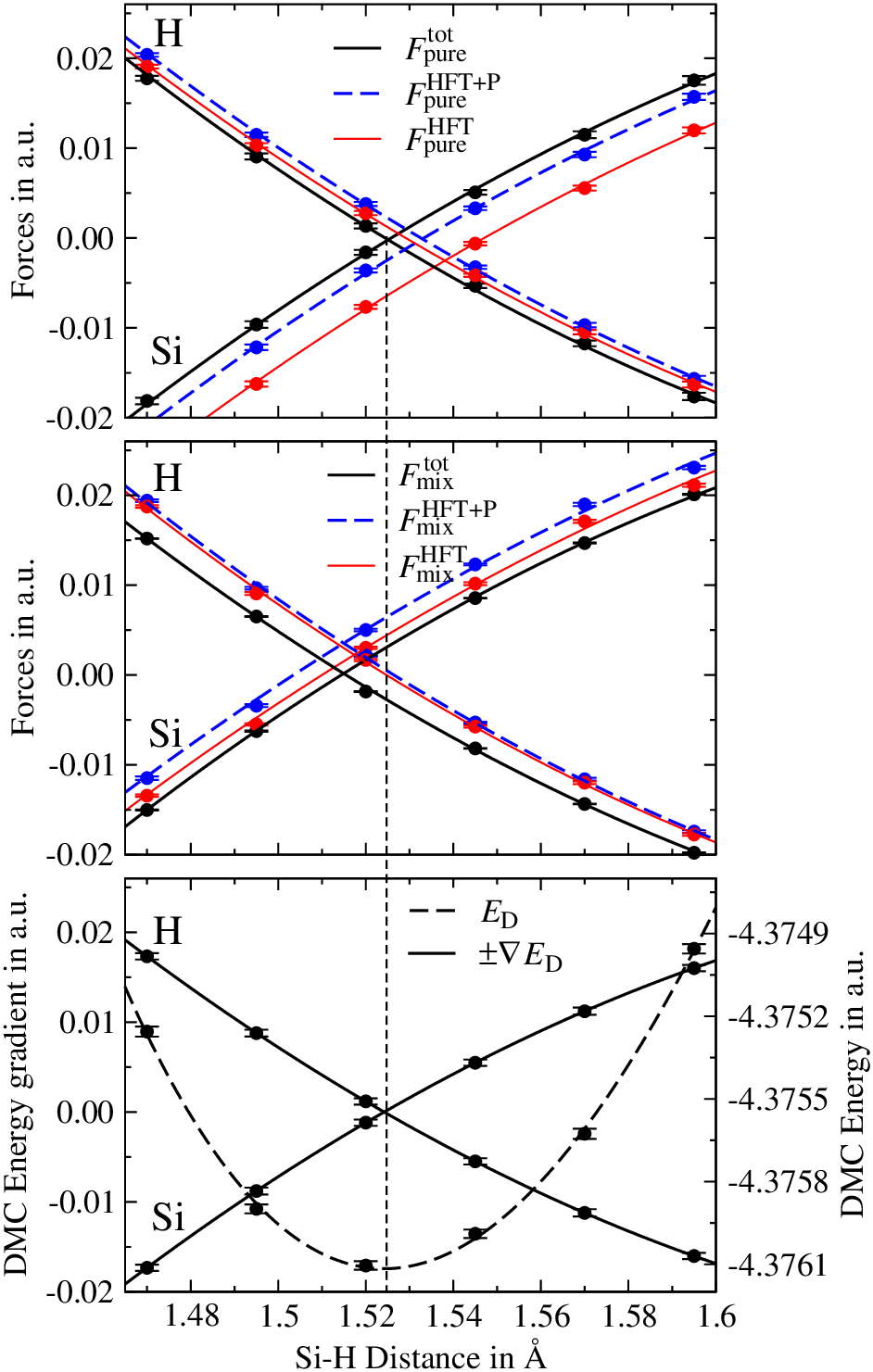}
\par\end{centering}
\caption{\label{fig:sih_example}(Color online) Upper graph: three
different estimates of the pure DMC forces,
$F_{\mathrm{pure}}^{\mathrm{HFT}}$,
$F_{\mathrm{pure}}^{\mathrm{HFT+P}}$ and
$F_{\mathrm{pure}}^{\mathrm{tot}}$, for the Si and H atoms of the SiH
molecule evaluated at different bond lengths. These estimates are
defined in Eqs.~(\ref{eq:F_tot_pure})-(\ref{eq:F_N_pure_approx}) and
are calculated with the small basis set. The Morse potential is fitted
to the forces. Middle graph: the same information as for the upper
graph with the three estimates, $F_{\mathrm{mix}}^{\mathrm{HFT}}$,
$F_{\mathrm{mix}}^{\mathrm{HFT+P}}$ and
$F_{\mathrm{mix}}^{\mathrm{tot}}$, calculated within the mixed DMC
method. These estimates are defined in
Eqs.~(\ref{eq:F_tot,mix})-(\ref{eq:Reynolds_approx}).  Lower graph:
the DMC energy at different bond lengths, a Morse potential fitted to
the DMC energies and the derivative of the Morse potential
corresponding to the forces on the Si and H atoms.  To guide the
reader, the dotted vertical line corresponds to the zero-force
geometry obtained from the fitted DMC energy curve.}
\end{figure}

For the SiH molecule, Figure \ref{fig:sih_example} shows different DMC
force estimates defined in
Eqs.~(\ref{eq:F_tot,mix})-(\ref{eq:F_N_pure_approx}) and evaluated at
different bond lengths. For reference, this figure also shows the DMC
energies which are simultaneously calculated with the forces. A Morse
potential fitted to the energies is also plotted together with its
gradient. The dotted vertical line indicates the zero-force geometry
obtained from the DMC energy curve. This figure shows that the
geometry obtained from the minimum of the potential energy curve
agrees well with the zero-force geometry obtained from the pure DMC
total force estimates.

Since we investigate molecules containing H atoms and a heavier atom,
one could obtain equilibrium bond lengths and vibrational frequencies
from the forces on the H atoms alone. However, to test the force
estimators on heavier atoms directly, we report bond lengths and
frequencies obtained using both the zero-force condition on the H atom
and on the heavier atoms. For SiH$_{4}$, the estimates of the forces
on the Si atom should be zero by symmetry and this condition is
satisfied within a standard error of less than 0.001\,a.u. Also, the
symmetries of the H$_{2}$ and SiH$_{4}$ molecules imply that the force
on each H atom should have the same magnitude. Since we found this to
hold within statistical errors, we average symmetry-related components
to further reduce the statistical error bar.

\section{Results\label{sec:Results}}

\subsection{Definitions}

We define the errors in quantities such as bond lengths and
frequencies as the differences between the values obtained from the
forces and from the energies,
\begin{equation} 
\Delta y_{\mathrm{method}}^{F}=y_{\mathrm{method}}^{F}-y^{E},
\label{eq:error}
\end{equation}
where $y$ is the bond length $a$ or vibrational frequency $\omega$,
and the ``method'' can be mixed or pure DMC. $E$ indicates the DMC
energy and $F$ stands for the different force estimators: HFT (the HFT
estimator), HFT+P (with the pseudopotential Pulay estimator) and
$F_{\mathrm{tot}}$, in the mixed DMC method as defined in
Eqs.~(\ref{eq:F_tot,mix})-(\ref{eq:Reynolds_approx}), and in the pure
DMC method as defined in
Eqs.~(\ref{eq:F_tot_pure})-(\ref{eq:F_N_pure_approx}).

In mixed DMC calculations, the error $\Delta y$ may arise from the
Reynolds' approximation of Eq.~(\ref{eq:Reynolds_approx}) and from
replacing all total derivatives with partial derivatives. Both
approximations introduce an error of first order in
$(\Psi_{\mathrm{T}}-\Phi)$. In pure DMC calculations, the error
$\Delta y$ may arise from the approximate nodal term in
Eq.~(\ref{eq:F_N_pure_approx}) and from replacing the total with
partial derivatives. The latter approximation introduces an error of
first order in ($\Psi_{\mathrm{T}}-\Phi$), whereas the approximate
form of the nodal term introduces an error of second order.

\subsection{Bond lengths}

\begin{table*}\begin{centering}\begin{tabular}{lllllll}
\hline\hline 
 & $a_{\mathrm{expt}}$& $a^{E}$& $a_{\mathrm{mix}}^{F^{\mathrm{tot}}}(1)$ & 
   $a_{\mathrm{pure}}^{F^{\mathrm{tot}}}(1)$ & 
   $a_{\mathrm{mix}}^{F^{\mathrm{tot}}}(2)$ & 
   $a_{\mathrm{pure}}^{F^{\mathrm{tot}}}(2)$\\
\hline
H$_2$  & 0.741& 0.7416(2)& 0.74090(3)& 0.7411(2)&  & \\
SiH    & 1.520& 1.5195(7)& 1.5123(1) & 1.5179(8)& 1.5107(1)& 1.5173(9)\\
GeH    & 1.589& 1.6012(8)& 1.5913(1) & 1.5993(9)& 1.5901(1)& 1.5992(10)\\
SiH$_4$& 1.480& 1.4740(4)& 1.46970(9)& 1.4728(10)&  &\\
\hline\hline 
\end{tabular}
\par\end{centering}
\caption{\label{tab:geometries}Equilibrium bond lengths
($\mathrm{\AA}$) for four molecules calculated from mixed and pure DMC
forces on each atom. The first column gives the experimental bond
lengths from Ref.~\citep{NIST}, the second column shows values
obtained from the DMC energy curves, and the other columns give bond
lengths obtained from the zero-force condition for the mixed and pure
DMC total force estimators as defined in Eqs.~(\ref{eq:F_tot,mix})
and~(\ref{eq:F_tot_pure}).  We specify bond lengths obtained from
forces on the H atoms by~(1) and on the non-H atoms by~(2). }
\end{table*}

\begin{table*}\begin{centering}\begin{tabular}{llcllllll}
\hline\hline 
& Basis & $a^{E}$ & 
$\,\,\,\Delta a_{\mathrm{mix}}^{\mathrm{HFT}}(1)$     & 
$\,\,\,\Delta a_{\mathrm{mix}}^{\mathrm{HFT+P}}(1)$   & 
$\,\,\,\Delta a_{\mathrm{mix}}^{F_{\mathrm{tot}}}(1)$ & 
$\,\,\,\Delta a_{\mathrm{pure}}^{\mathrm{HFT}}(1)$    & 
$\,\,\,\Delta a_{\mathrm{pure}}^{\mathrm{HFT+P}}(1)$  & 
$\,\,\,\Delta a_{\mathrm{pure}}^{F_{\mathrm{tot}}}(1)$\\
\hline
H$_2$  & large& 0.7416(2) & -0.0057(2)& -0.0057(2)&  -0.0007(2)& -0.0002(2) & -0.0003(2) & -0.0004(3)\\
SiH    & large& 1.5195(7) & -0.0098(8)& -0.0091(8)&  -0.0072(7)& -0.0006(10)& ~0.0007(10)& -0.0016(11)\\
GeH    & large& 1.6012(8) & -0.0138(9)& -0.0141(9)&  -0.0099(8)& -0.0012(11)& -0.0018(11)& -0.0019(12)\\
SiH$_4$& large& 1.4740(4) & -0.0055(6)& -0.0056(6)&  -0.0043(4)& -0.0005(8) & -0.0007(8) & -0.0012(11)\\
\hline 
SiH  &   small& 1.5242(6) & ~0.0006(7) & ~0.0023(7)&  -0.0091(6)& ~0.0052(8) & ~0.0086(8) & ~0.0004(9)\\
GeH  &   small& 1.5991(7) & -0.0059(7) & -0.0058(7)&  -0.0060(7)& ~0.0024(10)& ~0.0028(10)& -0.0010(11)\\
\hline\hline
&&&&&&&&\\
 & Basis & $a^{E}$ & 
 $\,\,\,\Delta a_{\mathrm{mix}}^{\mathrm{HFT}}(2)$ & 
 $\,\,\,\Delta a_{\mathrm{mix}}^{\mathrm{HFT+P}}(2)$ & 
 $\,\,\,\Delta a_{\mathrm{mix}}^{F_{\mathrm{tot}}}(2)$ & 
 $\,\,\,\Delta a_{\mathrm{pure}}^{\mathrm{HFT}}(2)$ & 
 $\,\,\,\Delta a_{\mathrm{pure}}^{\mathrm{HFT+P}}(2)$ & 
 $\,\,\,\Delta a_{\mathrm{pure}}^{F_{\mathrm{tot}}}(2)$\\
\hline
SiH   & large&  1.5195(7) & -0.0283(8)&  -0.0279(8)&  -0.0088(8)& -0.0032(10)& -0.0023(10)& -0.0022(11)\\
GeH   & large&  1.6012(8) & -0.0208(10)& -0.0200(10)& -0.0111(8)& -0.0044(12)& -0.0026(12)& -0.0020(13)\\
\hline 
SiH   & small&  1.5242(6) & -0.0142(7)&  -0.0207(7)& -0.0101(7)& ~0.0233(8)&   ~0.0090(8) & ~0.0016(9)\\
GeH   & small&  1.5991(7) & -0.0220(8)&  -0.0206(8)& -0.0085(7)& -0.0027(10)&  ~0.0001(10)& -0.0023(12)\\
\hline\hline 
\end{tabular}
\par\end{centering}
\caption{\label{tab:geometries_error}Differences $\Delta a$ in the
equilibrium bond lengths ($\mathrm{\AA}$) derived from the DMC forces
and from the DMC energies. $\Delta a$ is defined by
Eq.~(\ref{eq:error}).  The first column specifies the basis set used,
the second column gives bond lengths obtained from the DMC energy
curves, and the other columns give bond lengths obtained from the
zero-force condition for the three different estimators,
$F^{\mathrm{HFT}}$, $F^{\mathrm{HFT+P}}$, and $F^{\mathrm{tot}}$, in
the mixed and pure DMC methods, defined in
Eqs.~(\ref{eq:F_tot,mix})-(\ref{eq:F_N_pure_approx}).  We specify bond
lengths obtained from forces on the H atoms by~(1) and on the non-H
atoms by~(2). }
\end{table*}

Table \ref{tab:geometries} presents equilibrium bond lengths
calculated with the mixed and pure DMC total force estimators for four
molecules using the FPLA. Table \ref{tab:geometries_error} gives
further details of the different contributions to the total forces and
presents the difference, $\Delta a$, of the bond lengths derived from
the forces and from the energies, as defined in
Eq.~(\ref{eq:error}). 

We begin by discussing pure DMC results with the large basis set. The
HFT estimates $F_{\mathrm{pure}}^{\mathrm{HFT}}$ from forces on the H
atoms give very accurate bond lengths (upper part of Table
\ref{tab:geometries_error}).  As shown in the third last column of
Table \ref{tab:geometries_error}, the errors $\Delta
a_{\mathrm{pure}}^{\mathrm{HFT}}(1)$ in the bond lengths derived from
the forces on the H atoms are not much larger than one standard error
of 0.001\,\AA.  The bond lengths from the HFT forces calculated on the
non-H atoms (lower part of Table \ref{tab:geometries_error}) are not
as accurate.
Adding the pseudopotential Pulay term to the HFT forces,
$F_{\mathrm{pure}}^{\mathrm{HFT+P}}$, has a very small effect on the
bond lengths derived from the forces on the H atoms, as shown in the
penultimate column of Table \ref{tab:geometries_error}. For forces
acting on the non-H atoms, however, adding the pseudopotential Pulay
term improves the bond lengths slightly.  The nodal terms are very
small and do not significantly change the bond lengths obtained in the
large basis set pure DMC calculations.
The error bars on $\Delta a_{\mathrm{pure}}^{F_{\mathrm{tot}}}$ are
dominated by the contribution from the HFT force, so that including
the pseudopotential Pulay and nodal terms does not increase the noise
much.

Table \ref{tab:geometries_error} shows that both the pseudopotential
Pulay and nodal terms become more important for SiH and GeH when using
the small basis set.  The bond lengths from the HFT forces on both the
H and Si atoms are significantly worse than for the large basis set.
However, when the pseudopotential Pulay and nodal terms are included
the bond lengths are not significantly worse than for the large basis
set.

When comparing the mixed and pure DMC total force estimates, we find
from Table \ref{tab:geometries} that the statistical errors in all
bond lengths obtained from the mixed DMC forces are about a factor ten
smaller.  This is because the pure DMC estimator used here does not
satisfy a zero-variance condition.
The absolute deviation in all bond lengths derived from the mixed DMC
total forces and from the energies is on average 0.0076(2)\,\AA. In a
similar comparison, the pure DMC total forces give bond lengths with a
much smaller absolute average deviation of 0.0015(4)\,\AA. Although
adding the Pulay terms to the mixed DMC HFT force may improve the bond
lengths by up to 17 standard errors in our results, all pure DMC
forces (with and without Pulay terms) generally give more accurate
bond lengths than the best mixed DMC total force estimates.  This
difference in accuracy can be understood by recalling that the error
introduced in the mixed DMC force estimates is of first order in
($\Psi_{\mathrm{T}}-\Phi$) whereas the error in the pure DMC force
estimates is only of second order.  The additional first order error
from replacing the total derivatives by partial derivatives appears to
be small.

The differences between the DMC bond lengths (from either the DMC
energy or the pure DMC forces) and experimental data in Table
\ref{tab:geometries} are somewhat larger than the difference between
the bond lengths from the DMC energies and forces. This difference is
largest for GeH (0.010(1)-0.012(1)\,\AA), followed by SiH$_{4}$
(0.006(0)-0.007(1)\,\AA) and SiH (0.001(1)-0.003(1)\,\AA) and is
negligible for H$_{2}$. These bond length deviations from experiment
must largely arise from a combination of the fixed-node approximation,
the FPLA scheme, which slightly alters the pure DMC ground-state
distribution, and the pseudopotentials. The fixed-node approximation
could be improved by using more accurate trial wavefunctions.  It is
more difficult to develop better pseudopotentials, although including
core-polarization potentials on the Si and Ge atoms may also improve
the results \citep{shirley,lee2,maezono}.

\subsection{Vibrational Frequencies}

\begin{table}\begin{centering}\begin{tabular}{lllllll}
\hline\hline 
 & $\omega_{\mathrm{expt}}$ & $\omega^{E}$& 
   $\omega_{\mathrm{mix}}^{F_{\mathrm{tot}}}(1)$ & 
   $\omega_{\mathrm{pure}}^{F_{\mathrm{tot}}}(1)$ & 
   $\omega_{\mathrm{mix}}^{F_{\mathrm{tot}}}(2)$ & 
   $\omega_{\mathrm{pure}}^{F_{\mathrm{tot}}}(2)$\\
\hline
H$_2$    & 4401&    4420(16)& 4441(2)& 4403(15)&  &\\
SiH      & 2042&    2049(14)& 2075(2)& 2041(11)&  2080(2)& 2052(12)\\
GeH      & 1908(35)&1907(14)& 1944(1)& 1909(12)&  1949(2)& 1904(13)\\
SiH$_4$  & 2187&    2288(11)& 2324(2)& 2299(29)&  &\\
\hline\hline 
\end{tabular}
\par\end{centering}
\caption{\label{tab:frequencies}Harmonic vibrational frequencies
(cm$^{-1}$) calculated within the mixed and pure DMC methods. The
abbreviations are analogous to those in Table \ref{tab:geometries}.}
\end{table}

\begin{table*}\begin{centering}\begin{tabular}{llcllllll}
\hline\hline 
 & Basis & $\omega^{E}$ & 
 $\Delta\omega_{\mathrm{mix}}^{\mathrm{HFT}}(1)$ & 
 $\Delta\omega_{\mathrm{mix}}^{\mathrm{HFT+P}}(1)$ & 
 $\Delta\omega_{\mathrm{mix}}^{F_{\mathrm{tot}}}(1)$ & 
 $\Delta\omega_{\mathrm{pure}}^{\mathrm{HFT}}(1)$ & 
 $\Delta\omega_{\mathrm{pure}}^{\mathrm{HFT+P}}(1)$ & 
 $\Delta\omega_{\mathrm{pure}}^{F_{\mathrm{tot}}}(1)$\\
\hline
H$_2$  &large& 4420(16)& 128(18)& 128(18)& 21(16)& -10(19)& -10(19)& -16(22)\\
SiH    &large& 2049(14)&~~68(15)&~~70(15)& 26(13)& -10(16)&~~-7(16)&~~-7(18)\\
GeH    &large& 1907(14)&~~87(15)&~~89(15)& 37(14)&~~~1(16)&~~~4(16)&~~~2(18)\\
SiH$_4$&large& 2288(11)&~~71(16)&~~74(15)& 36(11)& ~25(23)& ~30(23)& ~11(31)\\
\hline 
SiH    &small& 1992(12)&~~71(13)&~~72(13)& 81(12)& ~16(14)& ~19(15)& ~46(16)\\
GeH    &small& 1905(10)&~~74(11)&~~74(11)& 29(10)& -25(13)& -25(13)&~~~0(15)\\
\hline\hline
 &  &  &  &  &  &  &  & \\
 & Basis & $\omega^{E}$ & 
 $\Delta\omega_{\mathrm{mix}}^{\mathrm{HFT}}(2)$ & 
 $\Delta a_{\mathrm{mix}}^{\mathrm{HFT+P}}(2)$ & 
 $\Delta a_{\mathrm{mix}}^{F_{\mathrm{tot}}}(2)$ & 
 $\Delta a_{\mathrm{pure}}^{\mathrm{HFT}}(2)$ & 
 $\Delta a_{\mathrm{pure}}^{\mathrm{HFT+P}}(2)$ & 
 $\Delta a_{\mathrm{pure}}^{F_{\mathrm{tot}}}(2)$\\
\hline
SiH   &large& 2049(14)& 103(16)& 100(17)& 31(14)& ~14(16)&~~~1(17)& ~3(18)\\
GeH   &large& 1907(14)&~~73(16)&~~68(16)& 41(14)&~~~2(17)& -10(17)& -4(19)\\
\hline 
SiH   &small& 1992(12)& 104(13)& 129(13)& 92(12)& -23(17)& ~27(15)& 53(16)\\
GeH   &small& 1905(10)&~~86(12)&~~85(12)& 38(10)& ~23(14)& ~17(14)& 20(16)\\
\hline\hline 
\end{tabular}
\par\end{centering}
\caption{\label{tab:frequencies_error}Difference $\Delta\omega$ in the
harmonic vibrational frequencies (cm$^{-1}$) derived from the DMC
forces and energies. The difference $\Delta\omega$ is defined by
Eq.~(\ref{eq:error}). The abbreviations are analogous to those in
Table \ref{tab:geometries_error}. }
\end{table*}

Table \ref{tab:frequencies} presents harmonic vibrational frequencies
calculated with the mixed and pure DMC total force estimators. Table
\ref{tab:frequencies_error} gives details of the different
contributions to the total forces and presents the differences,
$\Delta\omega$, of the frequencies derived from the forces and from
the energies, as defined in Eq.~(\ref{eq:error}).

Table \ref{tab:frequencies_error} shows that for all pure DMC force
estimators the difference in the vibrational frequencies derived from
the forces and the energies is comparable to or less than one standard
error of about $20\,\mathrm{cm}^{-1}$ with the exception of SiH.
The effect of the pseudopotential Pulay and nodal terms on the pure
DMC frequencies is small.  This may also be seen qualitatively in
Figure \ref{fig:sih_example}, where adding the Pulay terms mostly
shifts the forces at different bond lengths by similar amounts.

As in the discussion of bond lengths, we find that the mixed DMC total
forces give less accurate frequencies than the pure DMC forces. The
absolute difference between frequencies derived from the mixed DMC
total forces and from the energies is on average 43(4)\,cm$^{-1}$
compared to 16(6)\,cm$^{-1}$ when the frequencies are derived from
pure DMC total forces. Although adding the Pulay terms to the mixed
DMC HFT force may improve the frequencies by up to ten standard errors
in our results, all pure DMC force estimates (with and without Pulay
terms) still give more accurate frequencies than the best mixed DMC
total force estimates.

\subsection{Comparison of the FPLA and SPLA schemes}

\begin{table*}\begin{centering}\begin{tabular}{lccllcccc}
\hline\hline 
 & PLA & $a^{E}$ & 
 $\Delta a_{\mathrm{mix}}^{\mathrm{HFT}}(1)$ & 
 $\Delta a_{\mathrm{mix}}^{\mathrm{HFT+P}}(1)$ & 
 $\Delta a_{\mathrm{mix}}^{F_{\mathrm{tot}}}(1)$ & 
 $\Delta a_{\mathrm{pure}}^{\mathrm{HFT}}(1)$ & 
 $\Delta a_{\mathrm{pure}}^{\mathrm{HFT+P}}(1)$ & 
 $\Delta a_{\mathrm{pure}}^{F_{\mathrm{tot}}}(1)$\\
\hline
SiH   & FPLA& 1.5195(7)& -0.0098(8)& -0.0091(8)& -0.0072(7)& -0.0006(10)& ~0.0007(10)& -0.0016(11)\\
      & SPLA& 1.5210(8)& -0.0113(9)& -0.0110(9)& -0.0086(8)& -0.0024(10)& -0.0019(10)& -0.0034(11)\\
GeH   & FPLA& 1.6012(8)& -0.0138(9)& -0.0141(9)& -0.0099(8)& -0.0012(11)& -0.0018(11)& -0.0019(12)\\
      & SPLA& 1.5995(9)& -0.0129(10)&-0.0128(10)& -0.0073(9)&-0.0004(11)& -0.0004(11)& -0.0039(13)\\
\hline\hline
 &  &  &  &  &  &  &  & \\
& PLA & $a^{E}$ & 
$\Delta a_{\mathrm{mix}}^{\mathrm{HFT}}(2)$ & 
$\Delta a_{\mathrm{mix}}^{\mathrm{HFT+P}}(2)$ & 
$\Delta a_{\mathrm{mix}}^{F_{\mathrm{tot}}}(2)$ & 
$\Delta a_{\mathrm{pure}}^{\mathrm{HFT}}(2)$ & 
$\Delta a_{\mathrm{pure}}^{\mathrm{HFT+P}}(2)$ & 
$\Delta a_{\mathrm{pure}}^{F_{\mathrm{tot}}}(2)$\\
\hline
SiH    & FPLA & 1.5195(7)& -0.0283(8)&  -0.0279(8)&  -0.0088(8)& -0.0032(10)& -0.0023(10)& -0.0022(11)\\
       & SPLA & 1.5210(8)& -0.0303(9)&  -0.0289(9)&  -0.0098(8)& -0.0055(10)& -0.0026(10)& -0.0034(12)\\
GeH    & FPLA & 1.6012(8)& -0.0208(10)& -0.0200(10)& -0.0111(8)& -0.0044(12)& -0.0026(12)& -0.0020(13)\\
       & SPLA & 1.5995(9)& -0.0157(10)& -0.0148(10)& -0.0084(9)& -0.0060(12)& -0.0042(12)& -0.0033(15)\\
\hline\hline 
\end{tabular}
\par\end{centering}
\caption{\label{tab:geometries_PLA}Difference $\Delta a$ in the
equilibrium bond lengths ($\mathrm{\AA}$) derived from the DMC forces
and energies for SiH and GeH. The FPLA and SPLA localization
approximations are used as indicated in the first column. All other
abbreviations are the same as in Table \ref{tab:geometries_error}.}
\end{table*}

\begin{table*}\begin{centering}\begin{tabular}{lccllllll}
\hline\hline 
 & PLA  & $\omega^{E}$ & 
 $\Delta\omega_{\mathrm{mix}}^{\mathrm{HFT}}(1)$ & 
 $\Delta\omega_{\mathrm{mix}}^{\mathrm{HFT+P}}(1)$ & 
 $\Delta\omega_{\mathrm{mix}}^{F_{\mathrm{tot}}}(1)$ & 
 $\Delta\omega_{\mathrm{pure}}^{\mathrm{HFT}}(1)$ & 
 $\Delta\omega_{\mathrm{pure}}^{\mathrm{HFT+P}}(1)$ & 
 $\Delta\omega_{\mathrm{pure}}^{F_{\mathrm{tot}}}(1)$\\
\hline
SiH & FPLA & 2049(14)& 68(15)& 70(15)& 26(13)& -10(16)& ~~-7(16)& ~~-7(18)\\
    & SPLA & 2050(12)& 73(14)& 73(13)& 24(12)& ~13(15)&  ~11(15)& ~~~6(18)\\
GeH & FPLA & 1907(14)& 87(15)& 89(15)& 37(14)&~~~1(16)& ~~~4(16)& ~~~2(18)\\
    & SPLA & 1915(13)& 33(14)& 33(14)& 20(13)& -19(15)&  -19(15)&  ~20(18)\\
\hline\hline 
 &  &  &  &  &  &  &  & \\
& PLA &  $\omega^{E}$ & 
 $\Delta\omega_{\mathrm{mix}}^{\mathrm{HFT}}(2)$ & 
 $\Delta a_{\mathrm{mix}}^{\mathrm{HFT+P}}(2)$ & 
 $\Delta a_{\mathrm{mix}}^{F_{\mathrm{tot}}}(2)$ & 
 $\Delta a_{\mathrm{pure}}^{\mathrm{HFT}}(2)$ & 
 $\Delta a_{\mathrm{pure}}^{\mathrm{HFT+P}}(2)$ & 
 $\Delta a_{\mathrm{pure}}^{F_{\mathrm{tot}}}(2)$\\
\hline
SiH  & FPLA& 2049(14)& 103(16)&  100(17)& 31(14)& ~14(16)& ~~~1(17)&~3(18)\\
     & SPLA& 2050(12)&~~95(16)& ~~89(15)& 28(12)& ~19(15)& ~~~6(16)&~2(19)\\
GeH  & FPLA& 1907(14)&~~73(16)& ~~68(16)& 41(14)&~~~2(17)&  -10(17)&-4(19)\\
     & SPLA& 1915(13)&~~25(15)& ~~25(15)& 24(13)&~~~3(17)& ~~-3(16)&~2(20)\\
\hline\hline 
\end{tabular}
\par\end{centering}
\caption{\label{tab:frequencies_PLA}Difference $\Delta\omega$ in the
harmonic vibrational frequencies (cm$^{-1}$) derived from the DMC
forces and from the DMC energies for SiH and GeH. The FPLA and SPLA
localization approximations are used as indicated in the first
column. All other abbreviations are the same as in Table
\ref{tab:geometries_error}. }
\end{table*}

The FPLA and SPLA schemes are compared when calculating forces for the
SiH and GeH molecules. Since the schemes generate different
ground-state wavefunctions, expectation values may also differ between
the two schemes.  Additionally, we also use slightly different
approximations in the force estimators under the FPLA and SPLA
schemes. The calculated bond lengths and vibrational frequencies are
presented in Tables \ref{tab:geometries_PLA} and
\ref{tab:frequencies_PLA}. When bond lengths and frequencies are
derived from any of the three pure DMC force estimates, the results
between the two localization schemes agree within or close to one
standard error of about 0.0015\,\AA\ for the bond lengths and about
20\,cm$^{-1}$ for the frequencies. When the results are instead
derived from the mixed DMC forces, we find that the difference can be
as large as three standard errors.

\section{Conclusion}

We have presented exact expressions for the total atomic forces within
mixed and pure diffusion Monte Carlo (DMC) calculations using nonlocal
pseudopotentials and reported approximations for estimating
them. These expressions include the pseudopotential Pulay
term\citep{badinski1}
and the Pulay nodal term\citep{badinski2}.

We obtained harmonic vibrational frequencies and equilibrium bond
lengths from the calculated forces for four small molecules. The
calculations were performed with single-determinant Slater-Jastrow
trial wavefunctions using the mixed DMC and future-walking
\citep{Future} pure DMC methods. In the pure DMC calculations we found
that the contributions to the force from the Pulay nodal term and the
pseudopotential Pulay term were comparable to or less than the
statistical error in the total force, when the trial wavefunction and
the nodal surface were sufficiently accurate. In these cases,
neglecting the nodal and pseudopotential Pulay terms could have been
justified. However, when the trial wavefunctions were less accurate,
both Pulay terms became important and including them significantly
improved the pure DMC forces. All bond lengths and vibrational
frequencies derived from the pure DMC total forces agreed with those
obtained from the DMC energies within or close to one standard
error. This showed that the error from replacing total with partial
derivatives in the pure DMC force estimators is very small, and that
the additional error from approximating the pure DMC nodal term is
well behaved.

The deviations of the bond lengths and frequencies obtained from the
mixed DMC total forces and from the energies were generally much
larger than in the pure DMC calculations. This was explained by the
less accurate approximations in the mixed DMC force estimates which
introduce errors of first order in $(\Psi_{\mathrm{T}}-\Phi)$. For a
specified quality of trial wavefunction we therefore concluded that
pure DMC forces were more accurate than mixed DMC ones. We also
investigated both the FPLA and SPLA schemes for treating the nonlocal
pseudopotential operator and found very similar results.

A brief review of previous attempts to calculate forces within the DMC
method and a discussion of the performance of various quantum
chemistry methods in estimating bond lengths and vibrational
frequencies for several molecules was presented in
Ref.~\citep{badinski1}. The deviation between our results and
experimental data is comparable to or better than results obtained by
other accurate quantum chemistry methods, and is generally
considerably better than in density functional methods.  Our work has
demonstrated that accurate atomic forces can be calculated with
pseudopotentials and the DMC method.

\subsection*{Acknowledgements}

We are grateful to John Trail for helpful discussions.  This work was
supported by the Engineering and Physical Sciences Research Council
(EPSRC) of the United Kingdom. Computing resources were provided by
the University of Cambridge High Performance Computing Service (HPCS).

\bibliographystyle{apsrev}

\end{document}